\documentstyle[12pt,a4]{article}
\normalsize
\def\simless{\mathbin{\lower 3pt\hbox{$\rlap{\raise 5pt\hbox{$\char'074$}}
\mathchar"7218$}}}
\def\simgreat{\mathbin{\lower 3pt\hbox{$\rlap{\raise 5pt \hbox{$\char'076$}}
\mathchar"7218$}}}
\catcode`@=11

\def\beqra{\begin{eqnarray}} \def\eeqra{\end{eqnarray}}
\def\beq{\begin{equation}}      \def\eeq{\end{equation}}

\def\fo{\hbox{{1}\kern-.25em\hbox{l}}}

\def\ch{\@startsection{section}{1}{\z@}{-3ex plus-1ex minus-.2ex}%
        {2ex plus.2ex}{\large\sc}}

\def\; \lapp \;{\raisebox{-.4ex}{\rlap{$\sim$}} \raisebox{.4ex}{$<$}}

\def\con{\ifmmode \hbox{\bf*} \else{\bf*}\fi}   % conjugation
\def\scon{\ifmmode \hbox{\footnotesize\rm\bf*} \else{\footnotesize\rm\bf*}\fi}

\def\0#1{\relax\ifmmode\mathaccent"7017{#1}%    % puts a little circle atop,
        \else\accent23#1\relax\fi}              % as a halo of a saint

\def\eslash{\not{\hbox{\kern-2pt $E$}}}
\catcode`@=12

\begin{document}
\hoffset=0.4cm
\voffset=-1truecm
\normalsize

%%%%%%%%%%%%%%%%%%%%%%%%%%%%

\begin{titlepage}

\begin{flushright}
DFPD 93/TH/75\\
SISSA 93/183-A\\
\end{flushright}
\vspace{20pt}
\centerline{\Large {\bf Thermal Background Corrections}}
\vskip 0.3 cm
\centerline{\Large {\bf to the  Neutrino  Electromagnetic Vertex
}}
\vskip 0.3 cm
\centerline{\Large{\bf in Models with Charged
Scalar Bosons }}
\vspace{12pt}
\begin{center}
{\large\bf A. Riotto$^{a,b,}$\footnote{Email:riotto@tsmi19.sissa.it}}
\end{center}
\vskip 0.2 cm
\centerline{\it $^{(a)}$International School for Advanced Studies,
SISSA-ISAS}
\centerline{\it Strada Costiera 11, 34014 Miramare, Trieste, Italy}
\vskip 0.2 cm
\centerline{\it $^{(b)}$Istituto Nazionale di Fisica Nucleare,}
\centerline{\it Sezione di Padova, 35100 Padua, Italy.}
\vskip 0.2 cm

\baselineskip=20pt
\vskip 0.5 cm
\centerline{\large\bf Abstract}
\vskip 0.2 cm

\end{titlepage}
\centerline{{\large {\bf 1. Introduction}}}
\vskip 0.8 cm
\baselineskip=24pt
It is well-known that the properties of neutrinos propagating through a
medium differ form those in the vacuum; for instance, the vacuum
energy-momentum relation for massless neutrinos $E_\nu=\left|{\bf
p}_{\nu}\right|$, where $E_\nu$ is the energy and $\left|{\bf
p}_{\nu}\right|$ is the magnitude of the momentum vector, is no longer
valid in the medium \cite{wol}. The modification of the neutrino
dispersion relation can be represented in terms of an index of
refraction or an effective potential and arise, in the framework of
finite-temperature field theory, from the temperature- and
density-dependent corrections to the neutrino self-energy \cite{ref}.

Of primary interest along this line is also the study of the
electromagnetic interactions of neutrinos in a medium \cite{rad}. The
dramatic enhancement of the radiative decay rate of neutrinos in an
electron-rich medium has been investigated in ref. \cite{en} and is due
to the fact that the Glashow-Iliopoulos-Maiani (GIM) mechanism, which
suppresses the radiative decay in vacuum, is inoperative for the matter
contribution. Moreover, since the medium can introduce ${\cal CP}$ and
${\cal CPT}$ asymmetries in the effective potential interactions,
Majorana neutrinos are allowed to get diagonal electric and magnetic
dipole moments \cite{rad} in the Standard Model (SM)
which are forbidden in the vacuum.

The great attention in the recent literature on the properties of
neutrinos propagating in a medium has been motivated by the attractive
suggestion that the solar neutrino problem \cite{sun} can be solved by
the resonant oscillation mechanism \cite{os}. Another possible
explanation of the observed neutrino deficit from the sun is based on
the interactions of solar neutrinos with the magnetic field of the outer
layers of the sun. This requires large diagonal \cite{magn} and/or
transition \cite{marciano} magnetic moments for the electron neutrino,
of order of $\left(10^{-11}-10^{-10}\right)\:\mu_B$, where $\mu_B$ is
the Bohr magneton. Moreover, light neutrinos possessing a magnetic moment
of order of $10^{-12}\: \mu_B$ can play a role in many astrophysical
phenomena such as the rapid cooling of degenerate stars and (if
neutrinos are of the
Dirac type) the emission of the collapse energy from the core of supernovae
\cite{raf}. Also, a value of $\mu_\nu\sim 10^{-12}\:\mu_B$ is
cosmologically acceptable for Dirac neutrinos \cite{morg}.

Unfortunately in SM the magnetic
moment is generated at the one-loop level and is extremely small because
the only scales of the problems are the mass of the neutrino
$m_{\nu_{e}}$ and
the Fermi constant $G_F$. Indeed \cite{mom},
\beq
\mu_{\nu_{e}}=\frac{3\: e\: G_{F}}{8\:\pi^2\: \sqrt{2}}\: m_{\nu_{e}}\simeq
3\times 10^{-19}\:\left(\frac{m_{\nu_{e}}}{1\:\mbox{eV}}\right)\:\mu_B.
\eeq
It is clear that to get a
magnetic moment of order of $10^{-12}\:\mu_B$ one has to invoke
some new physics beyond the SM. Indeed, a large class of models
\cite{13,14} which are able to provide large magnetic
moments to neutrinos, have the common feature to posses new charged
scalar bosons whose mass can be arbitrary \cite{13} or fixed by the
supersymmetric scale \cite{14}.

In the present work we give the results of detailed calculations of the
background-dependent part of the $\nu\nu\gamma$ vertex
when these new charges scalar bosons
couple to leptons in a medium
consisting of a gas of electrons. As usual, the electron gas is embedded
in a uniform positive-ion background. However, the effect of ions is
negligible in most circumstances \cite{bra}.

For sake of concreteness, we have decided to perform all the calculations
in a well-defined framework, namely the supersymmetric model with
explicit breaking of $R$-parity \cite{14,rparity}, where neutrinos are
Majorana particles. The generalization to other models for both Majorana
and Dirac neutrinos is straightforward \cite{rif}.

We find that the magnetic (electric) dipole
moment does not receive from the medium
any significant enhancement, as suggested by Giunti {\it et al.} in
ref. [3] for the SM . However, a new chirality flipping,
but helicity conserving,
term is induced by the interactions with the thermal bath. This new term
vanishes if the background is ${\cal
CPT}$ symmetric and is associated to the longitudinal photon exchanged and
therefore
disappears in the vacuum. We estimate the contribution coming from this
new term to the plasmon decay process $\gamma_{pl}\rightarrow \nu\nu$
\cite{raf}, which is the primary source of the rapid cooling of
degenerate stars, and show that it can be comparable to the contribution
due to the vacuum magnetic moment.

We also show that, as in the case of SM \cite{rad}, one-loop thermal
corrections bring in an effective charge for Majorana neutrinos in a
medium as well as a magnetic (electric) diagonal
 dipole moment which would not be allowed in the vacuum. Moreover, the
effective
potential receives a correction in presence of an external magnetic
field.

The paper is organized as follows. In Section {\bf 2} we present the
model we have adopted to illustrate our calculations. In Section {\bf 3}
the calculations are described and general formulas for the form factors
are given in terms of integrals over the electron-positron
energy distribution. Some details of the calculations and the results in
different limits are given in the Appendix. Then in Section {\bf 4}
we estimate the plasmon decay rate contribution from the new terms
arising in the medium. Section {\bf 5} presents our conclusions.
\vspace{1 cm}
\\
\centerline{{\large \bf 2. The model}}

\vskip 0.8 cm

The minimal supersymmetric standard model \cite{nilles} with explicit
$R$-parity breaking \cite{rparity} via $L$-violation is described by the
superpotential which, in addition to the standard Yukawa couplings,
involves the $\Delta L\neq 0$ couplings
\beq
f^{\Delta L\neq 0}=\frac{1}{2}\lambda_{ijk}\:\left[L_i,L_j\right]e^{c}_k+
\lambda^{\prime}_{ijk}\:L_i Q_j d^{c}_{k},
\eeq
where, $i,j,k$ are generation indices, $L$, $Q$ are the lepton and the
quark left-handed doublets and $e^c$, $d^c$ are (the charge conjugate
of) the right-handed lepton and charge $-1/3$ quark singlets,
respectively. The first term in eq. (2) gives rise to the Lagrangian
\beq
{\cal L}^{\Delta L\neq 0}=\lambda_{ijk}\left[
\tilde{l}^{j}_{L}\bar{l}^{k}_{R}\nu^{i}_{L}+
\left(\tilde{l}_{R}^{k}\right)^{*}\left(\bar{\nu}^{i}_{L}\right)^c
l^{j}_{L}+\tilde{\nu}_{L}^{i}\bar{l}_{R}^{k}l_{L}^{j}-\left(
i\leftrightarrow j\right)\right]+\mbox{h.c.},
\eeq
where $l^c=C\bar{l}^T$ means the charge conjugated of the fermion $l$,
$C$ being the charge conjugation matrix,
and we have disregarded the second term in eq. (2) since we are
interested in a medium consisting of electrons and positrons.

In the vacuum the couplings of eq. (3) give rise to neutrino masses
and magnetic moments (after the insertion of a photon vertex in any
charged internal line) through two different one-loop diagrams, see
figure 1. In all the diagrams of figure 1 an helicity flip on the
internal fermion line is necessary. As explicitly indicated, this also
requires a mixing of the scalar leptons associated with the different
chiralities. Any of the diagrams of figure 1 contribute to
$m_{\nu_{i}\nu_{j}}$ and $\mu_{\nu_{i}\nu_{j}}$ as
\beqra
m_{\nu_{i}\nu_{j}}&\simeq&\frac{\lambda_{a}\lambda_{b}}{16\pi^2}
\:m\:\frac{{\rm
sin}2\theta_k}{2}\left(\frac{1}{m_{1k}^2}-\frac{1}{m_{2k}^2}\right),\\
\mu_{\nu_{i}\nu_{j}}&=&\mu_B\frac{\lambda_{a}\lambda_{b}}{8\pi^2}
\:m_{e}\:m\:{\rm sin}2\theta_{k}\left\{\frac{1}{m_{1k}^2}\left[
{\rm ln}\left(\frac{m_{1k}^2}{m^2}\right)-1\right]\right.\nonumber \\
&-&
\left.\frac{1}{m_{2k}^2}\left[
{\rm ln}\left(\frac{m_{2k}^2}{m^2}\right)-1\right]\right\},
\eeqra
where $m$ is the mass of the internal charged lepton, $\lambda_a$ and
$\lambda_b$ are the appropriate couplings, and $m_{1k}$, $m_{2k}$ and $
\theta_{k}$ are the two mass eigenvalues and the mixing angle
of the $\tilde{l}^{k}_L\tilde{l}^{k}_R$ mixing matrix, respectively.

If, for instance, we examine the $\nu_e\nu_\mu\gamma$ vertex, since the
contribution to $\mu_{\nu_e\nu_\mu}$ turns out to be proportional
to $m_{\nu_e\nu_\mu}$ and we require $m_{\nu_e\nu_\mu}<{\cal O}(10)
$ eV, a strong bound on $\mu_{\nu_e\nu_\mu}$ is obtained, roughly
$\mu_{\nu_e\nu_\mu}<{\cal O}(10^{-14})\mu_B$. To enhance the vacuum
magnetic moment $\mu_{\nu_e\nu_\mu}$ to the astrophysically
interesting value of $
10^{-12}\mu_B$, one can follow ref. \cite{14} and impose that the lepton
number $L_e-L_\mu$ remains unbroken and that the Lagrangian,
in the limit of vanishing Yukawa
couplings, is symmetric under an $SU(2)_{H}$ horizontal symmetry acting
on the first and the second generations. Under this assumption, the only
terms which survive in eq. (2) are
\beq
f^{\Delta L\neq 0}=\lambda_{123}L_eL_\mu \tau^c\:+\:
\lambda_{131}\left(L_eL_\tau e^c+L_\mu L_\tau \mu^c\right).
\eeq
The corresponding graphs giving rise to $m_{\nu_e\nu_\mu}$ and
$\mu_{\nu_e\nu_\mu}$ are given in figure 2. Since under the horizontal
symmetry $SU(2)_H$ the mass term $m_{\nu_e\nu_\mu}$ is odd, whereas
$\mu_{\nu_e\nu_\mu}$ is even, diagrams 2a) and 2b) and 2c) and 2d) tend
to cancel out and to sum up for $m_{\nu_e\nu_\mu}$ and
$\mu_{\nu_e\nu_\mu}$,
respectively. As a consequence, $\mu_{\nu_e\nu_\mu}$ is now no longer
proportional to $m_{\nu_e\nu_\mu}$ and the value
$\mu_{\nu_e\nu_\mu}\simeq
10^{-12}\mu_B$ can be achieved \cite{14}.

If we now consider a medium filled up with a gas of electrons, the
Lagrangian which gives rise to the finite-temperature effective
vertices involving $\nu_{e}$, $\nu_{\mu}$ and $\gamma$ is
\beq
{\cal L}=\lambda_{131}\tilde{\tau}_{L}\bar{e}_{R}\nu_{e L}\:-\:
\lambda_{123}\tilde{\tau}_{R}^{*}\left(\bar{\nu}_{\mu L}\right)^c e_L
\:+ \mbox{h.c.}.
\eeq
This will be our starting Lagrangian in the next Section\footnote{
Even if we are focusing on a particular vertex, $\nu_e\nu_\mu\gamma$, in
a particular model, we want to stress again that the structure of the
 form factors
derived in the next Section are model-independent.}.
\newpage
\centerline{{\large \bf 3. Calculation of the vertex functions in the
medium}}

\vskip 0.4 cm

\flushleft{\bf 3.1 Chirality flipping terms}
\vskip 0.8 cm

In this Subsection we calculate the contribution to the chirality
flipping term $\left(\nu_{\mu L}\right)^c\nu_{e L}\gamma$ in the medium
assuming that the temperature is such that there are no charged scalar
particles $\tilde{\tau}_L$ and $\tilde{\tau}_R$ in the background.
Therefore, only the electron propagator has a background-dependent
part and is given by
\beq
S_F(k)=\left(\not\!k +m_e\right)\left[\frac{1}{k^2+m_e^2}
+2\pi i\delta\left(k^2-m_e^2\right)\eta\left(k\cdot u\right)\right],
\eeq
where
\beq
\eta(x)=\frac{\theta(x)}{{\rm e}^{\beta\left(x-\mu\right)}+1}+
\frac{\theta(-x)}{{\rm e}^{-\beta\left(x-\mu\right)}+1}.
\eeq
Here $\theta(x)$ is the unit step function, $1/\beta$ is the
temperature,
$\mu$ is the electron chemical potential and $u^\mu$ is the
four-velocity of the center of mass of the medium.

The off-shell electromagnetic
vertex function $\Gamma_\mu^{LR}\left(p_1,p_2,u\right)$ is
defined in such a way
that
\beq
\langle\nu_{\mu}\left(p_1\right)\left|J_{\mu}^{EM}(0)\right|
\nu_e\left(p_2\right)\rangle\equiv\bar{u}_{\mu}\left(p_1\right) \Gamma_\mu^{LR}
\left(p_1,p_2,
u\right) u_{e}\left(p\right).
\eeq
Note that in the vacuum the dependence on $u^\mu$ vanishes. The
diagrams which enter the calculation of $\Gamma_{\mu}^{LR}$ are shown in
figure 3.

Since the integrals involved in the calculations of $\Gamma_{\mu}^{LR}$
are cut off by the electron-positron distribution, the diagram 3b) gives
a contribution to $\Gamma_{\mu}^{LR}$ suppressed by an extra power of
$1/\tilde{m}^2$, $\tilde{m}$ being the typical supersymmetric mass,
relative to the diagram 3a). Therefore, we neglect it.

With this preliminaries, we have to calculate the following
quantity
\beqra
-i{\cal G}^{LR}_\mu&=& e\frac{\lambda_{123}\lambda_{131}}{2}{\rm sin}2\theta_3
\:\int\frac{d^4
k}{\left(2\pi\right)^4}\:iS_F\left(k-q\right)\gamma_\mu
iS_{F}\left(k\right)L\nonumber\\
&\times&
\left[\frac{1}{\left(k-p\right)^2-m_{23}^2}-\frac{1}{\left(k-p\right)^2-
m_{13}^2}\right],
\eeqra
where $L=\left(1-\gamma_5\right)/2$ is the left-handed
chirality operator.

When the electron propagator from eq. (8) is plugged in eq. (11),
several terms are produced beyond the standard vacuum term. The
terms with two factors $\eta(k\cdot u)$ contribute only to the
absorptive part of the amplitude (see, for instance, D'Olivo {\it et
al.} in ref. \cite{rad} for further comments on this points). In this paper we
will calculate only the dispersive part of the form factors. We also
make the local approximation, {\it i.e.} neglect the momentum dependence
of the heavy charged scalar bosons; hence ${\cal G}_\mu^{LR}$ reduces to
\beqra
{\cal G}_{\mu}^{LR}&=& e\frac{\lambda_{123}\lambda_{131}}{2}{\rm sin}2\theta_3
\left[\frac{1}{m_{23}^2}-\frac{1}{m_{13}^2}\right]
\:\int\frac{d^4
k}{\left(2\pi\right)^3}\left(\not\!k-\not\!q+m_e\right)\gamma_{\mu}
\left(\not\!k+m_e\right)\nonumber\\
&\times& \left\{\frac{\delta\left[\left(k-q\right)^2-m_e^2\right]}{
k^2-m_e^2}\eta\left[\left(k-q\right)\cdot u\right]\right.\nonumber\\
&+&\left.
\frac{\delta\left(k^2-m_e^2\right)}{\left(
k-q\right)^2-m_e^2}\eta\left(k\cdot u\right)\right\}L.
\eeqra
Making the change of variable $k\rightarrow k+q$ in the first integral
of eq. (12) we obtain
\beqra
{\cal G}_{\mu}^{LR}&=& e\frac{\lambda_{123}\lambda_{131}}{2}{\rm sin}2\theta_3
\left[\frac{1}{m_{23}^2}-\frac{1}{m_{13}^2}\right]\:m_e\nonumber\\
&\times&\left({\cal I}_{\mu}^1+{\cal I}_{\mu}^2+{\cal I}_{\mu}^3
\right)L,
\eeqra
where we have defined
\beqra
{\cal I}_{\mu}^1&=&\int\frac{d^3
k}{\left(2\pi\right)^3}\frac{\left(f_{-}-f_{+}\right)}{2E}
\frac{4 k^\mu q^2}{\left(q^2+2 k\cdot q\right)\left(q^2-2 k \cdot
q\right)},\\
{\cal I}_{\mu}^2&=&\int\frac{d^3
k}{\left(2\pi\right)^3}\frac{\left(f_{-}+f_{+}\right)}{2E}
\frac{-2i\sigma_{\mu\nu}q^\nu\:q^2}
{\left(q^2+2 k\cdot q\right)\left(q^2-2 k \cdot
q\right)},\\
{\cal I}_{\mu}^3&=&\int\frac{d^3
k}{\left(2\pi\right)^3}\frac{\left(f_{-}-f_{+}\right)}{2E}
\frac{-4 \left(k\cdot q\right)\: q_\mu}
{\left(q^2+2 k\cdot q\right)\left(q^2-2 k \cdot
q\right)},
\eeqra
and $\sigma_{\mu\nu}=(i/2)\left[\gamma_{\mu},\gamma_{\nu}\right]$,
$k^\mu=\left(E,{\bf k}\right)$, $E=\sqrt{\left|{\bf
k}\right|+m_e^2}$, and we have introduced the electron and positron
distributions
\beq
f\mp\left(k\right)=\frac{1}{{\rm e}^{\beta\left(k\cdot u\mp
\mu\right)+1}}.
\eeq
Note that $q^\mu {\cal G}_\mu^{LR}=0$ due to the electromagnetic gauge
invariance.

If the initial Lagrangian (in the vacuum)
of our model respects ${\cal
CP}$ (so that we take all the $\lambda$'s real), we can define the
four-component self-conjugate states
$\chi_a=\nu_a+\eta_a\left(\nu_a\right)^c$, where $\eta_a=\pm 1$ are
 the intrinsic
${\cal CP}$ parities of $\chi_a$'s. It is well-known that if $\chi_e$ and
$\chi_\mu$ have opposite (equal) ${\cal CP}$ parities, then their off-diagonal
dipole (magnetic) moment vanishes \cite{nieves}. Moreover, the correct
expression for $\Gamma_{\mu}^{LR, Majorana}$ can be derived form expressions
(13-16) once one remembers that $\chi_e$ and $\chi_\mu$ are Majorana
neutrinos, so that for each Feynman diagram there exists a second
diagram in which all the internal particles are replaced by their charge
conjugates \cite{nieves}. A practical rule to derive
$\Gamma_{\mu}^{LR,Majorana}$
is to treat neutrinos as Dirac particles and then  add to
$\Gamma_{\mu}^{LR,Dirac}$ its charge conjugate part
\beqra
\Gamma_{\mu}^{LR,Majorana}\left(p_1,p_2,u\right)&=&
\Gamma_{\mu}^{LR,Dirac}\left(p_1,p_2,u\right)\nonumber\\
&+&\eta_e\eta_\mu\:\gamma^0\left[
C\:\Gamma_{\mu}^{LR,Dirac}\left(-p_1,-p_2,u\right)\:C^{-1}\right]^{*}\gamma^0.
\eeqra

If we now introduce the following tensors
\beqra
\tilde{g}_{\mu\nu}&=&g_{\mu\nu}-\frac{q_{\mu}q_{\nu}}{q^2},\\
\tilde{u}_\mu&=& \tilde{g}_{\mu\nu} u^\nu,
\eeqra
and remind that, as a consequence of having taken the limit
$\tilde{m}\rightarrow\infty$, the form factors  depend only on $q$ and
not on
$p_1$ and $p_2$ separately,
the complete off-shell electromagnetic vertex function  reads
\beq
\Gamma_{\mu}^{LR,Majorana}=
\left[F_1\tilde{u}_{\mu}\left(L+\eta_e\eta_\mu R\right)+
i\frac{F_2}{2}\left(1-\eta_e\eta_\mu\right) \sigma_{\mu\nu}q^\nu+
i\frac{F_2}{2}\sigma_{\mu\nu}\gamma_5\left(1+\eta_e\eta_\mu\right)
q^\nu\right],
\eeq
where $R=\left(1+\gamma_5\right)/2$ is the right-handed chirality
operator and
\beqra
F_1&=& e\frac{\lambda_{123}\lambda_{131}}{2}{\rm sin}2\theta_3
\left[\frac{1}{m_{23}^2}-\frac{1}{m_{13}^2}\right]\:m_e
\:2q^2\nonumber\\
&\times& \int\frac{d^3
k}{\left(2\pi\right)^2}\frac{\left(f_{-}-f_{+}\right)}{2E}\frac{\left(u\cdot
k\right)}{\left(q^2+2 k\cdot q\right)\left(q^2-2 k \cdot
q\right)},\\
F_2&=&-e\frac{\lambda_{123}\lambda_{131}}{2}{\rm sin}2\theta_3
\left[\frac{1}{m_{23}^2}-\frac{1}{m_{13}^2}\right]\:m_e
2q^2\nonumber\\
&\times& \int\frac{d^3
k}{\left(2\pi\right)^2}\frac{\left(f_{-}+f_{+}\right)}{2E}
\frac{1}{\left(q^2+2 k\cdot q\right)\left(q^2-2 k \cdot
q\right)}.
\eeqra

The convenience of presenting the form factors $F_1$ and $F_2$ in this
form relies on the fact that $F_1$ and $F_2$ are scalars
so that the integrations can be performed in
the rest-frame of the medium defined by setting
$u^\mu=\left(1,\bf{0}\right)$. In th rest frame of the medium, we will denote
the components of the four vector $q^\mu$  by
\beq
q^\mu=\left(\Omega,\mbox{\boldmath ${\cal Q}$}\right),
\eeq
where $\Omega\equiv q\cdot u$ and
$\left|\mbox{\boldmath ${\cal Q}$}\right|\equiv\sqrt{\Omega^2-q^2}$
are manifestly scalar functions.

{}From expression (21) it is clear that the form factor $F_2$ can be
regarded as an additional contribution to the magnetic (or electric)
dipole moment. We note that, since
the contribution to the magnetic (electric) dipole moment in the medium
must be coherent with the neutrino propagation, it is necessary to take
the limit $q^\mu\rightarrow 0$ in eq. (23). Depending on how one
approaches the limit, eq. (23) can yield
different results because of its divergent nature. However, when
$q^\mu=0$, the internal electron lines with four-momenta $k$ and $k-q$
are on the mass shell, {\it i.e.} in the limit  $q^\mu\rightarrow 0$,
the diagram 3a) describe the process $\nu_e e\rightarrow \nu_\mu e$
with a modification of the external electron lines by the electromagnetic
field. Therefore one can expect  no enhancement in the medium
for the magnetic (electric) dipole moment (see Giunti {\it et al.} in
ref. [3]). Our result confirms this expectation.

The existence of the flipping term proportional
to $F_1$ is a unique property of the thermal bath. It is due to the
presence of the longitudinal photons which can couple to the internal
electron without changing its helicity. Let us recall that in the
vacuum, the magnetic (electric) moment flips both chirality and helicity
(at the leading order), since the photon is purely transverse, while the
vertex $F_1\tilde{u}^\mu$ cannot change the helicity of the incoming
neutrino. Moreover, if the chemical potential of the electron background
is zero, then $f_{+}=f_{-}$ and, therefore, $F_1=0$ for both Dirac and
Majorana neutrinos. Indeed, if we start from a ${\cal CP}$ invariant
Lagrangian in the vacuum and, for $\mu=0$, the background is symmetric,
$F_1$ must satisfy the relation
\beq
F_1\left(-p_1,-p_2,u\right)=-F_1\left(p_1,p_2,u\right),
\eeq
independently of whether the neutrino is of the Dirac type or Majorana
type. Since in our case $F_1$ is only a function of $p_1-p_2$, eq.
(25) implies that it is zero. However, we have ${\cal CP}$ as well as $
{\cal CPT}$ asymmetries in the medium due to
the presence of a nonvanishing electron chemical potential
 and therefore $F_1$ is not
zero in general\footnote{For a detailed discussion on the
electromagnetic properties of neutrinos in a medium and the role played by
discrete symmetries, see the first ref. in [3].}. Note also that, in the
particular model we have adopted to perform our calculations, the
$F_1$ term does not respect the horizontal $SU(2)_H$ symmetry
of the starting Lagrangian (7) and it can be nonvanishing only because
the medium, filled up with electrons and not with muons, breaks this
symmetry.

The integrals (22) and (23) can be worked out analytically in different
regimes either for a soft photon, namely when the exchanged momentum $q$
is much smaller than the momenta of the thermalized electrons (of order
of $T$ or $\mu$), or for
a hard, but almost light-cone photon, $q^2\rightarrow 0$.
The
calculations for the different
regimes, like the  ultrarelativistic and the classical ones,
can be found in the Appendix. We report here
only the result for a degenerate electron gas $\left(T\ll
\mu-m_e\right)$ since they are of interest for the calculations of the
plasmon decay rate which will be performed in the next Section.  We obtain
\beqra
F_1&=& e\frac{\lambda_{123}\lambda_{131}}{2}{\rm sin}2\theta_3
\left[\frac{1}{m_{23}^2}-\frac{1}{m_{13}^2}\right]\:m_e\:
\frac{1}{16\pi^2}\frac{q^2}{m_e^2\Omega^2}\frac{k_F^3}{3},\\
F_2&=&e\frac{\lambda_{123}\lambda_{131}}{2}{\rm sin}2\theta_3
\left[\frac{1}{m_{23}^2}-\frac{1}{m_{13}^2}\right]\:m_e\:
\frac{1}{4\pi^2}\frac{q^2}{m_e^2\Omega^2}\nonumber\\
&\times&\left[\frac{k_F E_F}{2}+m_e^2\:{\ln }\left(
\frac{k_F+E_F}{m_e}\right)\right],
\eeqra
where $E_F=\sqrt{m_e^2+k_F^2}=\mu$ is the Fermi energy.

\vskip 0.4cm
\flushleft{\bf 3.2 Chirality conserving terms}
\vskip 0.8cm

In this Subsection we calculate the contribution from the medium to the
chirality conserving term $\nu_e\nu_e\gamma$ (an analogous calculation
can be performed for other vertices).

Repeating the same considerations of the Subsection {\bf 3.1}, we have
to calculate the quantity (see figure 4)
\beqra
{\cal G}_{\mu}^{LL}&=& e\left|\lambda_{131}\right|^2
\left[\frac{{\rm cos}^2\theta_3}{m_{23}^2}+\frac{{\rm
sin}^2\theta_3}{m_{13}^2}
\right]\:\int\frac{d^4
k}{\left(2\pi\right)^3}\delta\left(k^2-m_e^2\right)\eta\left(k\cdot
u\right)\nonumber\\
&\times&\left\{\frac{4q^2\not\!k k_\mu+q^2\left[-2i\sigma_{\mu\nu}q^\nu-2k_\mu
\not\!q+2\left(k\cdot q\right)\gamma_\mu\right]}{
\left(q^2+2 k\cdot q\right)\left(q^2-2 k \cdot
q\right)}\right.\nonumber\\
&+&\left.\frac{-2\left(k\cdot q\right)
\left[2\not\!k q_\mu-2k_\mu\not\!q +2\left(k\cdot
q\right)\gamma_\mu\right]}{\left(q^2+2 k\cdot q\right)\left(q^2-2 k \cdot
q\right)}\right\}L.
\eeqra
Again $q^\mu {\cal G}_{\mu}^{LL}=0$ for the electromagnetic gauge
invariance.

It is then easy to show that the complete electromagnetic vertex function
$\Gamma_{\mu}^{LL}$ can be written under the form
\beqra
\Gamma_{\mu}^{LL}&=&\left[\tilde{F}_1\tilde{u}_\mu \not\!u+i
\tilde{F}_2\sigma_{\mu\nu}q^\nu\not\!u\right.\nonumber\\
&+&\left. i\tilde{F}_3\left(\gamma_{\mu}u_\nu-\gamma_\nu
u_\mu\right)q^\nu\not\!u+\tilde{F}_4 \tilde{g}_{\mu\nu}\gamma^\nu\right)L,
\eeqra
where
\beqra
\tilde{F}_1&=&-\frac{4}{3}e\left|\lambda_{131}\right|^2
\left[\frac{{\rm cos}^2\theta_3}{m_{23}^2}+\frac{{\rm
sin}^2\theta_3}{m_{13}^2}
\right]\:q^2\int\frac{d^3k}{\left(2\pi\right)^3}\frac{\left(f_{-}+f_{+}\right)}{
2E}\nonumber\\
&\times&\frac{\left[4\left(k\cdot u\right)^2-m_e^2\right]}{
\left(q^2+2 k\cdot q\right)\left(q^2-2 k \cdot
q\right)},\\
\tilde{F}_2&=& F_1,\\
\tilde{F}_3&=& i\:q^2 \left(q\cdot u\right) \tilde{F}_1,\\
\tilde{F}_4&=&\frac{8}{3}
e\left|\lambda_{131}\right|^2
\left[\frac{{\rm cos}^2\theta_3}{m_{23}^2}+\frac{{\rm
sin}^2\theta_3}{m_{13}^2}
\right]\:q^2\int\frac{d^3k}{\left(2\pi\right)^3}\frac{\left(f_{-}+f_{+}\right)}{
2E}\nonumber\\
&\times&\frac{\left[\left(k\cdot u\right)^2-m_e^2\right]}{
\left(q^2+2 k\cdot q\right)\left(q^2-2 k \cdot
q\right)}.
\eeqra
The expression (29) holds if neutrinos are of the Dirac type. If they
are of the Majorana type, then, applying the relation (18),
 $\Gamma_{\mu}^{LL,Majorana}$ can be easily obtained by adding
 to
$\Gamma_{\mu}^{LL}$ its self conjugate term. We then obtain
\beqra
\Gamma_{\mu}^{LL,Majorana}&=&-\left[\tilde{F}_1\tilde{u}_\mu \not\!u-i
\tilde{F}_2\sigma_{\mu\nu}q^\nu\not\!u\gamma_5 \right.\nonumber\\
&+&\left. i\tilde{F}_3\left(\gamma_{\mu}u_\nu-\gamma_\nu
u_\mu\right)q^\nu +\tilde{F}_4 \tilde{g}_{\mu\nu}\gamma^\nu\right)\gamma_5,
\eeqra
where we have make use of the property $\gamma_5^2={\bf 1}$.

The expressions for the form factors in the different regimes can be
found in the Appendix.

The physical interpretation of the form factors can be obtained
considering an interaction with an external field. Thus, taking the
external field of the form $A^\mu=\left(\phi,{\bf 0}\right)$ in the rest
frame of the medium, we see that $\tilde{F}_4$ yields an additional
contribution to the charge radius; moreover $\tilde{F}_3$ can be
regarded as an additional contribution to the electric dipole moment
and $\tilde{F}_2$ to the magnetic one. It is well known that Majorana
neutrinos can have neither a charge radius nor
 diagonal electric or magnetic dipole moments in the vacuum
\cite{nieves} since, for instance, $\bar{\nu}\sigma_{\mu\nu}\nu=0$ for
Majorana neutrinos. Nevertheless, already in the SM they can have electric or
magnetic dipole moments in a medium of electrons which introduces ${\cal
CP}$ as well ${\cal CPT}$ asymmetries \cite{rad}. We have found that
additional contributions can be given by some new physics beyond the SM.
These new contributions are again nonvanishing only if the medium
is ${\cal
CP}$ and ${\cal CPT}$ asymmetric. Let's take, for instance, the
contribution to the magnetic moment. In the non relativistic limit it
reduces to
\beq
{\cal O}_{M}=\tilde{F}_2\left(\bf{s}+\bar{{\bf s}}\right)\cdot{\cal {\bf
B}},
\eeq
where $\bf{s}$ and $\bar{\bf{s}}$ are
the spin expectation values for the particles and antiparticles and
${\cal {\bf B}}$ represents a uniform magnetic field. ${\cal O}_M$ is odd under
both ${\cal C}$ and ${\cal CPT}$. Since there are strong theoretical
reasons to believe that ${\cal CPT}$ is conserved by the Lagrangian in
the vacuum, any breaking of ${\cal CPT}$ must come from the background.
Indeed, the particles of the medium must have some chemical
potentials associated with them, otherwise $\tilde{F}_2$ must vanish.

We now consider the scattering of an electron neutrino with an external
static and uniform magnetic field ${\cal {\bf B}}$. The Dirac equation for a
neutrino spinor $\psi_\nu$ in the medium can be written as\footnote{
Even in the SM the effective potential receive a correction in presence
of a static and uniform magnetic field, see D'Olivo {\it et al.}
 in ref. \cite{rad}, but not proportional to $\not\!u$ as in the class
of models considered here.}
\beq
{\cal V}\psi_\nu=\left[\left(1-a_L\right)\not\!k+\left(b_L+c_L\right)
\not\!u\right]\psi_\nu=0,
\eeq
where ${\cal V}$ is called effective potential. The coefficients
$a_L$ and $b_L$ have been calculated in \cite{ref} whereas
\beqra
c_L&=&\mbox{\boldmath $\mu$}\cdot{\cal {\bf B}},\nonumber\\
\mbox{\boldmath $\mu$}&=&i\tilde{F}_2\left(\Omega=0,\left|
\mbox{\boldmath ${\cal Q}$}\right|\rightarrow 0\right)
\mbox{\boldmath $\sigma$}.
\eeqra
The value of $\tilde{F}_2$ in the limit $\Omega=0$, $\left|
\mbox{\boldmath ${\cal Q}$}\right|\rightarrow
0$ has been found, in the non relativistic limit, by D'Olivo {\it et
al.} in ref. \cite{rad}  and reads
\beq
\tilde{F}_2\left(\Omega=0,\left|\mbox{\boldmath ${\cal Q}$}
\right|\rightarrow 0\right)=\frac{\beta
n_{-}}{4 m_e},
\eeq
where $n_{-}={\rm
e}^{\beta\left(\mu-m_e\right)}\left(2m_e/\beta\right)^{3/2}\left[\Gamma\left(3/2
\right)/2\pi^2\right]$ is the number density of electrons.

The meaning of eq. (36) is that, in presence of a magnetic field, the
effective potential of a neutrino propagating through a medium gets a
new contribution proportional to $\left|{\cal {\bf B}}\right|$. The relative
importance of the matter density effects thus depend on the magnitude
of ${\cal {\bf B}}$. For instance, in the sun $\left|{\cal {\bf
B}}\right|$ is a few tenth of Tesla, the temperature is of order of 1
KeV, so that $c_L/b_L$ is very small,
$c_L/b_L\simeq \left|\lambda_{131}\right|^2 10^{-11}$ for
$m_{23}\simeq m_{13}\simeq$ 100 GeV. Nevertheless, application to other
physical contexts of this new term $c_L$ remains an open question and
should be kept in mind.
\vspace{1 cm}
\\
\centerline{{\large \bf 4. Plasmon decay}}

\vskip 0.8 cm
It is well known that in a medium composed by a gas of electrons the
dispersion relations for transverse and longitudinal photons are quite
different from those in the vacuum \cite{raf}. Indeed, both modes,
called plasmons,
acquire an effective plasma mass which allow them  to decay into a pair
of neutrinos.
The process $\gamma_{pl}\rightarrow \nu\nu$ represents the primary
source for the energy loss of degenerate plasmas, such as red giants and
white dwarfs \cite{raf} and has recently received considerable attention
\cite{bra}. It is well known that already in the SM plasmons can decay
into a pair of neutrinos \cite{raf}. If neutrinos couple to the photons
through a magnetic (or electric)
moment in the vacuum, the rate of the energy loss of
stellar plasmas due to $\mu_\nu$ is comparable to the SM contribution
for $\mu_\nu\simeq 10^{-12}\mu_B$ and no larger values of $\mu_\nu$ are
tolerated.

In this Section we want to estimate the contribution to the energy loss
through the plasmon decay induced by the chirality flipping term
proportional to $F_1$.

The differential decay rate of the process
$\gamma_{pl}(q)\rightarrow\nu_e\left(p_1\right)\nu_\mu\left(p_2\right)$ due to
the $F_1$ term in the rest frame of the medium (there is no
interference contribution with the magnetic moment term) is
\beq
d\Gamma_{F_1}=\frac{1}{2\Omega}\left(2\pi\right)^4\delta^{(4)}
\left(q-p_1-p_2\right)
\left|\bar{{\cal M}}_{F_1}\right|^2\frac{d^3 p_1}{\left(2\pi\right)^3 2E_1}
\frac{d^3 p_2}{\left(2\pi\right)^3 2E_2},
\eeq
where
\beqra
\left|\bar{{\cal M}}_{F_1}\right|^2&=& 4\left(
e\frac{\lambda_{123}\lambda_{131}}{2}{\rm sin}2\theta_3
\left[\frac{1}{m_{23}^2}-\frac{1}{m_{13}^2}\right]\:m_e
\frac{1}{2\pi^2}\frac{q^2}{m_e^2\Omega^2}\frac{k_F^3}{3}\right)^2\nonumber\\
&\times& \frac{\left(u\cdot
\eta_3\right)^2}{\left(\Omega\partial\varepsilon_L/\partial\Omega\right)}
\left(p_1\cdot p_2\right).
\eeqra
We have neglected neutrino masses and made use if the longitudinal
photon vector
\beq
\eta_3=\frac{1}{\left(q^2\right)^{1/2}}\left(\left|
\mbox{\boldmath ${\cal Q}$}\right|
,0,0,\Omega\right),
\eeq
which satisfies the relation $\eta_3\cdot q=0$.

In expression (40) $\varepsilon_L$ represents the dielectric constant
of the longitudinal plasmons \cite{raf} and, since we are considering
the case of degenerate stars, we are using the expression (26).

Using the Lenard's formula
\beq
\int\frac{d^3 p_1}{2 E_1}\frac{d^3 p_2}{2 E_2}
\delta^{(4)}\left(q-p_1-p_2\right)p_1^\mu p_2^\nu=
\frac{\pi}{24}\left(2p_1^\mu p_2^\nu+g^{\mu\nu} q^2\right),
\eeq
we find that
\beqra
\Gamma_{F_1}&=&\frac{1}{8\pi}\frac{1}{\Omega^2}\frac{\left|
\mbox{\boldmath ${\cal
Q}$}\right|^2}{\partial\varepsilon_L/\partial\Omega}\nonumber\\
&\times&
\left(
e\frac{\lambda_{123}\lambda_{131}}{2}{\rm sin}2\theta_3
\left[\frac{1}{m_{23}^2}-\frac{1}{m_{13}^2}\right]\:m_e
\frac{1}{2\pi^2}\frac{q^2}{m_e^2\Omega^2}\frac{k_F^3}{3}\right)^2.
\eeqra
The energy loss rate associated to the $F_1$ term is then given by
\beq
Q_{F_1}=\int\frac{d^3\mbox{\boldmath ${\cal Q}$}
}{\left(2\pi\right)^3}\frac{\Omega\: \Gamma_{F_1}}{
{\rm e}^{\Omega/T}-1}\left(\varepsilon_L-1\right)^2.
\eeq
In the case of white dwarfs (red giants before the helium flush) the
electron gas is degenerate with a temperature of order of (0.01-0.1) KeV
($\sim$ 8.6 KeV) and the Fermi momentum $k_F$ of order of 495 (400) KeV.
Therefore, rigorously speaking, the electron plasma is neither in the
nonrelativistic regime nor in the ultrarelativistic one \cite{bra}. In
such a case the expressions for $\varepsilon_L$ and the dispersion
relation for longitudinal photons are quite complicated and $Q_{F_1}$
can only be found numerically. Nevertheless, to have an idea of the
order of magnitude of $Q_{F_1}$, we can approximate $\varepsilon_L$ and
the dispersion relation as
\beqra
\varepsilon_L&=&1-\frac{\Omega_0^2}{\Omega^2}\left[1+\frac{3}{5}v_F^2
\frac{\left|\mbox{\boldmath ${\cal Q}$}\right|^2}{\Omega^2}\right],\\
q^2&=&\Omega_0^2+\left|{\bf{\cal
Q}}\right|^2\left[\frac{3}{5}v_{F}^2\frac{\Omega_0^2}{\Omega^2}-1\right],
\eeqra
where $\Omega_0=\left(4\alpha k_F^2 v_F/3\pi\right)^{1/2}$ is the plasma
frequency of order 10 KeV for both white dwarfs and red giants and
$v_F\sim 0.7$ is the Fermi velocity. Note that, since $\Omega_0\ll \mu$,
the expression (26) of $F_1$ valid for both soft and hard, but almost
light-cone photons, is a good approximation for the physical context
under consideration.

With such approximations, $Q_{F_1}$ can be expressed analytically and we
find
\beq
Q_{F_1}\simeq \frac{3}{64\pi^3}\frac{\Omega_0^5}{v_F^5}\left(
\frac{5}{3}\right)^{5/2}\frac{\sqrt{2\pi}}{\gamma^{5/2}}{\rm
 e}^{-\gamma}\: {\cal K}^2,
\eeq
where $\gamma=\Omega_0/T$ and we have expressed the coupling constants
in terms of the magnetic dipole moment $\mu_{\nu_e\nu_\mu}$
\beqra
{\cal K}&=& 4\left(\frac{{\rm sin}2\theta_3}{{\rm sin}2\theta_2}\right)
\left(\frac{1}{m_{13}^2}-\frac{1}{m_{23}^2}\right)\frac{k_F^3
\mu_{\nu_e\nu_\mu}}{6 m_e m_\tau}\nonumber\\
&\times&\left\{\frac{1}{m_{12}^2}\left[
{\rm ln}\left(\frac{m_{12}^2}{m^2}\right)-1\right]-
\frac{1}{m_{22}^2}\left[
{\rm ln}\left(\frac{m_{22}^2}{m^2}\right)-1\right]\right\}^{-1}.
\eeqra
In the last expression we have used the fact that the major contribution
to $\mu_{\nu_e\nu_\mu}$ comes from the diagram 2a) and 2b). In a similar
way one can find the energy loss rate due to the decay
$\gamma_{pl}\rightarrow\nu_e\nu_\mu$ through the magnetic dipole moment.
In the range of interest of temperatures and densities the longitudinal
and transverse contributions are comparable and, for instance,
\beq
Q_{long}\simeq\frac{1}{12}\frac{\mu_{\nu_e\nu_\mu}^2}{\left(2\pi\right)^3}
\left(\frac{5}{3}\right)^{3/2}\frac{\Omega_0^7}{v_F^3}\sqrt{2\pi}
\gamma^{-3/2}{\rm e}^{-\gamma}.
\eeq
The ratio between $Q_{F_1}$ and $Q_{long}$ is then
\beqra
\frac{Q_{F_1}}{Q_{long}}&\simeq&\frac{15}{2}\frac{1}{\mu_{\nu_e\nu_\mu}^2}
\frac{{\cal K}^2}{v_F^2\:\Omega_0^2\:\gamma}\nonumber\\
&\simeq&
10^{-5}\left(\frac{0.7}{v_F}\right)^2\left(\frac{10\:\mbox{KeV}}
{\Omega_0}\right)^2\left(\frac{10}{\gamma}\right)\left(\frac{k_F}{
400\:\mbox{KeV}}\right)^6\nonumber\\
&\times&
\left(\frac{{\rm sin}2\theta_3}{{\rm sin}2\theta_2}\frac{m_{13}^2-
m_{23}^2}{m_{12}^2-m_{22}^2}\right)^2\left(\frac{m_2}{m_3}\right)^8,
\eeqra
where we have indicated with $m_k$ the averaged eigenvalue of the mixing
matrix $\tilde{l}^{k}_{L}\tilde{l}_{R}^{k}$. Since in supersymmetric
models the factor
$\left({\rm sin}2\theta_3/{\rm sin}2\theta_2\right)\left(
m_{13}^2-
m_{23}^2/m_{12}^2-m_{22}^2\right)^2$ is of order of
$\left(m_\tau/m_\mu\right)$ \cite{nilles}, one can obtain,
$Q_{F_1}\simeq Q_{long}$ for $m_2\simeq 2 m_3$. Even if the above
estimation is approximate and holds in the particular framework we have
chosen, the general message one can read from it is that, going beyond
the SM, one has to take into account all the possible terms which arise
at finite temperature and density for the $\nu\nu\gamma$ vertex because
the new terms can give non negligible contributions to relevant
processes as the plasmon decay in degenerate stars.
\vspace{1 cm}
\\
\centerline{{\large \bf 5. Conclusions}}

In the present work we have carried out an explicit calculation of the
neutrino electromagnetic vertex in a background of electrons in a large
class of models where charged scalar bosons couple to leptons. We have
been motivated by the fact that such models are able to provide a
magnetic moment as large as $\mu_\nu\sim 10^{-12}\:\mu_B$, which can
play a relevant role in different astrophysical phenomena.

We have shown that the contribution from the medium to the magnetic
(electric) dipole moment is not significant, but a new chirality
flipping, but helicity conserving term, arises. This new term is
associated to the longitudinal photons and therefore disappears in the
vacuum and can be nonvanishing only because the medium does not respect
${\cal CPT}$. We have also estimated the contribution of this new term
to the plasmon decay rate showing that it can be comparable with the
contribution coming from the vacuum magnetic moment. Therefore it must
be taken into account in different applications of the vertex
$\nu\nu\gamma$ in a medium. Finally, we have calculated the correction
to the effective potential of a propagating neutrino in presence
of a magnetic field. Although the application of this to the solar
neutrino puzzle seems to be uninteresting, the possible applications in
other contexts deserve further consideration and are currently under
study.

\vspace{1. cm}

\centerline{\bf Acknowledgments}
\vspace{0.3 cm}
It is a pleasure to thank K. Enqvist and A. Masiero for reading the
early version of the paper and for useful suggestions.
\newpage
\setcounter{secnumdepth}{0}
\section{Appendix.}
%\centerline{\Large\bf Appendix.}
\setcounter{section}{1}
\renewcommand{\thesection}{\Alph{section}}
\renewcommand{\theequation}{\thesection . \arabic{equation}}
\vspace{.5 truecm}
\setcounter{equation}{0}

In this Appendix we give some details and the complete results for the
form factors introduced in the text for different regimes in the limit
of soft photons, namely when the exchanged momentum $q$ is much smaller
than the momenta of the thermalized electrons, or for a hard, but almost
light-like photon, $q^2\rightarrow 0$.
\vskip 0.4cm
\centerline{\bf Chirality flipping terms}
\vskip 0.4cm
\flushleft{\bf Ultrarelativistic regime $\left(T,\mu\gg m_e\right)$}
\vskip 0.4cm
\beqra
F_1&=&e\frac{\lambda_{123}\lambda_{131}}{2}{\rm sin}2\theta_3\left[
\frac{1}{m_{13}^2}-\frac{1}{m_{23}^2}\right]\nonumber\\
&\times&\frac{1}{8\pi^2 \beta}\left[a\left(m_e\beta,-\mu\right)-
a\left(m_e\beta,+\mu\right)\right],
\eeqra
where
\beq
a\left(m_e\beta,\pm\mu\right)={\rm ln}\left[1+{\rm e}^{-\left(
m_e\pm\mu\right)\beta}\right],\:\:\:\:\:T>\mu,
\eeq
and
\beq
a\left(m_e\beta,-\mu\right)\simeq\mu-m_e,\:\:\:\:\: T<\mu;
\eeq
\beqra
F_2&=&-e\frac{\lambda_{123}\lambda_{131}}{2}{\rm sin}2\theta_3\left[
\frac{1}{m_{13}^2}-\frac{1}{m_{23}^2}\right]\nonumber\\
&\times&\frac{1}{4\pi^2}\left[b\left(m_e\beta,-\mu\right)+
b\left(m_e\beta,+\mu\right)\right],
\eeqra
where
\beq
b\left(m_e\beta,\pm\mu\right)=\sum_{n=1}^{\infty}(-1)^n{\rm e}^{
\mp n\beta\mu}{\rm Ei}\left(-n\beta m_e\right),\:\:\:\:\:\: T>\mu,
\eeq
and
\beq
b\left(m_e\beta,-\mu\right)={\rm
ln}\left(\mu/m_e\right),\:\:\:\:\:\:\:T<\mu,
\eeq
where ${\rm Ei}(x)$ is the exponential-integral function.
\vskip 0.4cm
\flushleft{\bf Classical limit $(T\ll m_e$ and $(m_e-\mu)\gg
T)$}
\vskip 0.4cm
\beqra
F_1&=&-\frac{1}{8}e\frac{\lambda_{123}\lambda_{131}}{2}{\rm sin}2\theta_3\left[
\frac{1}{m_{13}^2}-\frac{1}{m_{23}^2}\right]\nonumber\\
&\times&q^2\frac{\sqrt{\pi}}{\Gamma(3/2)}\frac{n_{-}}{m_e^2\Omega^2},\\
F_2&=&\frac{1}{4}e\frac{\lambda_{123}\lambda_{131}}{2}{\rm sin}2\theta_3\left[
\frac{1}{m_{13}^2}-\frac{1}{m_{23}^2}\right]\nonumber\\
&\times&q^2\frac{\sqrt{\pi}}{\Gamma(3/2)}\frac{\beta n_{-}}{m_e^2\Omega^2},
\eeqra
where $n_{-}={\rm
e}^{\beta\left(\mu-m_e\right)}\left(2m_e/\beta\right)^{3/2}\left[\Gamma\left(3/2
\right)/2\pi^2\right]$.
\vskip 0.4cm
\centerline{\bf Chirality conserving terms}
\vskip 0.4cm

To calculate the chirality conserving form factors, we must calculate
the following integral
\beq
I_{\lambda\nu}=\int\frac{d^4 k}{\left(2\pi\right)^3}\frac{
\delta\left(k^2-m_e^2\right)\eta\left(k\cdot u\right)}{
\left(q^2+2q\cdot k\right)\left(q^2-2q\cdot k\right)}
k_{\lambda}k_{\nu}.
\eeq
Since $I_{\lambda\nu}=I_{\nu\lambda}$ and $I_{\lambda\nu}(q)=I_{\nu\lambda}
(-q)$, $I_{\lambda\nu}$ must be of the form
\beq
I_{\lambda\nu}=Au_{\lambda}u_{\nu}+B g_{\lambda\nu}+C
q_{\lambda}q_{\nu}.
\eeq
If we then contract $I_{\lambda\nu}$ with $u^{\lambda}$ the result must
be proportional to $u_{\nu}$, from which we read that $C=0$. From eq.
(A.9) we have
\beqra
I_1&=&u_{\lambda}u_{\nu}I^{\lambda\nu}=A+B=
\int\frac{d^4 k}{\left(2\pi\right)^3}\frac{
\delta\left(k^2-m_e^2\right)\eta\left(k\cdot u\right)\left(k\cdot
u\right)^2}{
\left(q^2+2q\cdot k\right)\left(q^2-2q\cdot k\right)},\\
I_2&=&g_{\lambda\nu}I^{\lambda\nu}=A+4B=
\int\frac{d^4 k}{\left(2\pi\right)^3}\frac{
\delta\left(k^2-m_e^2\right)\eta\left(k\cdot u\right)m_e^2}{
\left(q^2+2q\cdot k\right)\left(q^2-2q\cdot k\right)}.
\eeqra
The chirality conserving form factors are then functions of $I_1$ and
$I_2$
\beqra
\tilde{F}_1&=&\frac{1}{3}\left(4I_1-I_2\right),\nonumber\\
\tilde{F}_2&=&-2F_1,\nonumber\\
\tilde{F}_3&=&-i q^2\tilde{F}_1,\nonumber\\
\tilde{F}_4&=&\frac{1}{3}\left(I_2-I_1\right).
\eeqra
\vskip 0.4cm
\flushleft{\bf Ultrarelativistic regime $\left(T,\mu\gg m_e\right)$}
\vskip 0.4cm
\beqra
I_1&=&\frac{-1}{16\pi^2}\frac{1}{q^2\beta^2}\left[c\left(m_e\beta,+\mu\right)
+c\left(m_e\beta,-\mu\right)\right],\\
I_2&=&\frac{-m_e^2}{16\pi^2 q^2}\left[b\left(m_e\beta,+\mu\right)
+b\left(m_e\beta,-\mu\right)\right],
\eeqra
where
\beq
c\left(m_e\beta,\pm\mu\right)=\sum_{n=1}^{\infty}\frac{(-1)^n}{
n^2}{\rm e}^{-n\beta\left(m\pm\mu\right)},\:\:\:\:\:T>\mu,
\eeq
and
\beq
c\left(m_e\beta,-\mu\right)=\frac{\mu^2}{2},\:\:\:\:\:\:\: T<\mu.
\eeq
\vskip 0.4cm
\flushleft{\bf Degenerate limit $(T\ll m_e$ and $(\mu-m_e)\gg
T)$}
\vskip 0.4cm
\beqra
I_1&=&-\frac{1}{16\pi^2}\frac{1}{\Omega^2 m_e^2}\left[
\frac{E_F^3 k_F}{4}-\frac{E_F m_e k_F}{8}-\frac{
m_{e}^3}{8}{\rm ln}\left(\frac{E_F+k_F}{m_e}\right)\right],\\
I_2&=&-\frac{1}{16\pi^2}\frac{1}{\Omega^2}\left[
\frac{E_F k_F}{2}-\frac{
m_{e}^2}{2}{\rm ln}\left(\frac{E_F+k_F}{m_e}\right)\right].
\eeqra
\vskip 0.8cm
\flushleft{\bf Classical limit $(T\ll m_e$ and $(m_e-\mu)\gg
T)$}
\vskip 0.4cm
\beq
I_{1}\simeq I_2=-\frac{1}{16}\frac{1}{m_e^2\Omega^2}\frac{n_{-}
\sqrt{\pi}}{\Gamma(3/2)}.
\eeq

\newpage

%%%%%%%%%%%%%%%%%%--- References
%%%%%%%%%%%%%%%%%%%%%%%%%%%%%%%%%%%%%%%%%%%%%%%%%%%%%%%
\def\MPL #1 #2 #3 {Mod.~Phys.~Lett.~{\bf#1}\ (#3) #2}
\def\NPB #1 #2 #3 {Nucl.~Phys.~{\bf#1}\ (#3) #2}
\def\PLB #1 #2 #3 {Phys.~Lett.~{\bf#1}\ (#3) #2}
\def\PR #1 #2 #3 {Phys.~Rep.~{\bf#1}\ (#3) #2}
\def\PRD #1 #2 #3 {Phys.~Rev.~{\bf#1}\ (#3) #2}
\def\PRL #1 #2 #3 {Phys.~Rev.~Lett.~{\bf#1}\ (#3) #2}
\def\RMP #1 #2 #3 {Rev.~Mod.~Phys.~{\bf#1}\ (#3) #2}
\def\ZP #1 #2 #3 {Z.~Phys.~{\bf#1}\ (#3) #2}

%%%%%%%%%%%%%%%%%%%%%%%--- figures
\newpage
\noindent{\bf Figure Caption}
\begin{itemize}
\item[{\bf Fig. 1}]{Feynman diagrams contributing to $\mu_{\nu_i\nu_j}$
after a photon insertion line in any charged internal line.}
\item[{\bf Fig. 2}]{Feynman diagrams contributiong to $\mu_{\nu_e\nu_\mu}$
(after a photon insertion line in any charged internal line) when
$L_e-L_\mu$ conservation is imposed.}
\item[{\bf Fig. 3}]{Relevant Feynman diagram for the
$\nu_e\nu_\mu\gamma$ vertex in a background of electrons.}
\item[{\bf Fig. 4}]{Relevant Feynman diagram for the
$\nu_e\nu_e\gamma$ vertex in a background of electrons.}
\end{itemize}

\end{document}